**PAPER • OPEN ACCESS**

# Research on Intelligent Charging System Technology of Automobile Group

To cite this article: Kedi Yan 2019 *J. Phys.: Conf. Ser.* **1187** 022010

View the article online for updates and enhancements.

## Recent citations
- Cao LuCui

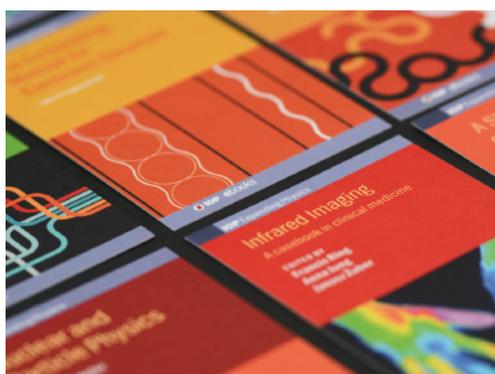

# Research on Intelligent Charging System Technology of Automobile Group

Kedi Yan

Oregon State University
97331

zzyankedi@gmail.com

**Abstract:** This paper analyzes the smart charging system for dealing with issues related to large parking garages, and analyzes the relevant technical standards of intelligent charging piles application and comprehensive transportation hubs. It mainly includes the number, area and charging method of the charging piles installed in the garages. New forms of construction management is conceived, and the investment and construction are reinforced, apportioning all charging system property rights to investment parties. Simultaneously, the investors take full responsibility for the future management and operation, mainly including the collection of service fees and charging fees.

**1. Intelligent charging pile technology**

*1.1 Alternating current (AC) charging*
As for this technology, it generally relies on three-phase AC power to provide the required power for electric vehicles. Since the vehicle-mounted charger and heat dissipation may affect it, the power of charging is small, and the time taken for charging is relatively long. The circuit configuration of a car charger usually has two types. If the correction circuit is added with the power factor in the previous stage, the influence on the harmonics of the power grid will be reduced, and the power factor state is still high.

*1.2 Direct current (DC) charging*
For this technology, it usually depends on the circuit composition such as control protection and the rectification filter. Among them, only three-phase uncontrollable and pwm-type rectification will utilize rectification technology. The three-phase uncontrollable rectification is relatively low in terms of capital, but the harmonic current is relatively large, requiring the equipment for harmonic control to be installed. The PWM type of rectified power has the characteristics of high efficiency and low harmonic current in conversion, but the capital requirement is relatively high.

*1.3 Wireless charging technology*

*1.3.1 Electromagnetic induction type*
This technology relies on two mutual inductance coils to complete wireless charging. When the input current of the coil changes, the magnetic field of the coils on the output side also changes, thereby generating an induced current. In this way, energy is passed to the output end through the input end. This technique of electromagnetic induction has no a long transport distance, but its energy conversion rate is relatively high.

*1.3.2 Magnetic field resonance type*
Similar to the principle of acoustic resonance, if the resonance frequencies of the two media are the same, energy transfer can be achieved. This type of wireless charging can also achieve simultaneous charging for a large number of devices, but the loss will be relatively high. Two schemes can be considered together to solve the problem of high charging loss (low efficiency) of this technology.

First, a feedback circuit is added at the transmitting end, which is: "parity-time-symmetric circuit incorporating a nonlinear gain saturation element". This component can automatically select the operating frequency according to the transmission distance to maximize the output power [6]. Second, metamaterials are utilized. Metamaterials benefit from its ability to focus magnetic flux, which can increase transmission efficiency and transmission distance. The combination of the two schemes may enable wireless charging to bring a new solution like spring wing to the pure electric vehicle charging problem.

## 2. Key component nodes analysis

*2.1 Charging control box of automobile group*
It is a type of outdoor equipment which is designed to integrates charging and power distribution devices and then installed into a sealed, moisture-proof, and rust-proof control box. This type of equipment is characterized by less occupied area and portfolio diversification. The integration of AC and DC can not only achieve AC charging and DC charging at the same time, but also make AC and DC arbitrarily combined to meet various needs. Meanwhile, real-time background monitoring on the charging process can be performed, so once an abnormality is found, the charging process can be automatically adjusted to achieve real-time protection. Real-time communication is maintained with the cloud platform, effectively protecting the charging vehicle and the power grid, so that the electricity utilization is guaranteed.

*2.2 Charging terminals*

*2.2.1 Single-phase AC terminal with a gun*
It is equipped with the control box and the charging box to provide the necessary charging interfaces, which is widely used in single-phase AC charging. At the same time, there must be a place where the charging gun is used, and according to its application, the area can be divided into the types of car bumper and wall-mounted. Among them, the wall-mounted type can be installed on the wall surface, with the requirement to avoid the rear end of the charging vehicle during installation. The single-phase AC wall-mounted charging terminal can be installed together with the bracket. For the type of car bumper, it can be used in underground garages and the place where the terrain is high and water is not easy to accumulate [1].

*2.2.2 Single-phase AC terminal with seats*
It cooperates with a charging port and a control box to provide a corresponding charging interface, mainly used in single-phase AC charging work, and the application place must be equipped with a charging socket.

Its main application is in single-phase AC charging, and it must be equipped with a charging socket. It is divided into two types: wall-mounted and floor-standing according to the application area.

*2.2.3 DC terminal with a gun*
It cooperates with a charging box to provide a corresponding charging interface. It is mainly used in DC

charging and must be equipped with a charging gun. According to the application area, it can be divided into two types of floor and wall-mounted. Among them, the floor type can be classified as single and double-sided in accordance with the site, while based on the different currents it can also be divided into two types of 125A and 250A.

**3. Intelligent charging system design scheme**

*3.1 System configuration scheme*
Due to the large difference in charging speed between DC charging and AC charging, a battery electric vehicle (with ordinary battery capacity) needs 7-9 h to be fully charged through the AC charging pile after being fully discharged, while it only requires 2-3 hours to be fully charged through the DC quick charging pile. Therefore, the AC and DC terminals are scientifically matched depending on the big data survey to achieve effective energy utilization and battery protection. For example, in the underground parkings of shopping malls, considering that most people will not park their cars for a long time, the combination ratio of DC and AC terminals can be decided to 8:2. As for the underground parking of the residential areas, taekeing into consideration that most people will park here overnight, the ratio of the DC terminals to AC terminals can chose 2:8. Taking a real project as an example, it is equipped with 80 channels of AC and 160 channels of DC terminals, which can allow more than 40 cars to carry out AC slow charging, simultaneously allowing over 160 cars to conduct DC fast charging. The charging function can be turned on by using the mobile phone software to scan the software QR code. In addition, the termination of charging can also be done by means of mobile phone softwares. The overall charging process can be queried by a mobile phone, with a high security. In the next 5-10 years, when the wireless charging is mature, due to its safe, stable and space-saving characteristics, it will gradually replace the AC charging devices, and even supersede the DC charging device if the efficiency problem is improved. The specific configuration scheme can be considered by referring to the following chart.

*3.2 Connection scheme*
The access power supply of the project is a 0.4kV low-voltage power supply, which needs to be introduced and connected in the substation. For low-voltage power access points, cables need to be installed between the charging devices of one car, which is done by the supplier [2]. The supplier shall formulate a scientific and reasonable engineering installation scheme according to the specific situation of the installation site. This can effectively ensure that the construction work meets the requirements and standards of electric power technology.

*3.3 Construction of the cloud platform*
It needs to monitor the operational power station information, the charging records and the status of terminal operations in real time. Meanwhile, all power station and terminal information data should be tracked to effectively analyze the data [3]. The cloud platform can not only effectively monitor the voltage, current and charge of the vehicle, but also efficaciously monitor the battery status of the vehicle with BMS.

*3.4 Security consideration*
For intelligent AC chargers, it needs to provide low-voltage power to the car, and it must have the protection to overvoltage, overload and over-temperature. While for the off-board vehicle DC charger, it is required to select the form of the component design. Among them, the electrical and business parts are required to be independent.

*3.5 Intelligent and secure billing system construction*
In terms of the charging control unit, the hardware and software interface need to be connected with the input and output components, so that the functions of switching control, data decryption and encryption, billing, human-computer display, etc. of the charging device can be finally realized.

### 4. Advantages of intelligent charging system

*4.1 Pile-free charging*
The difference from the conventional charging devices is that the pile-free charging does not need to rely on the charging piles. It transfers charging control and human-computer interaction to the charging box and the cloud respectively, with only one charging terminal in the local charging connection. However, the terminal is resistant to crushing, water repellency, and takes up less space. The pile-free charging transfers the human-computer interaction to the cloud platform, relying on the mobile APP and the cloud platform. As a result, the contact between people and public devices is reduced, making them more personalized and humanized. The pile-free charging can concentrate the control processes to share the communication and power modules, thus reducing the capital investment in the devices. At the same time, it can also make the control module to avoid some damage from the outside world and improve its security.

*4.2 Charging system structure module*
According to the structure of the product module and the prefabricated production equipment, the site only needs to build up the facility foundation [4]. The construction cycle of a small site is 7 days, and that of a large site is one month. For the intelligent charging technology, its update is extremely fast, and the modular structure is completed very quickly in terms of technology expansion and upgrade, making it consistent with the update of charging technology.

*4.3 Intelligent charging*
By means of centralized charging, wind energy, solar energy and battery energy storage technology can be effectively combined. In virtue of the function of the vehicle-mounted battery and the charging network, an intelligent microgrid system is established, and the system can operate independently in the grid through the grid-connected operation of the grid. Relying on the operation of the micro-scheduling of the power grid, it will not compete with people for electricity. It will use low-peak electricity to improve the efficiency of the grid. Flexible charging is usually realized on the basis of battery mechanism and characteristics. The charging device is used to determine the current and voltage according to the battery related requirements and the specific conditions, thereby restoring the function of the battery and achieving the purpose of life protection [5]. With the support of flexible charging technology, the traditional form of relying on BMS to determine the charging power is replaced, so that the charging safety is improved and the occurrence of safety accidents is avoided. Based on the collected battery related information, the data on the flexible charging is further determined, and the functions of life protection and battery recovery are realized in charging.

### 5. Conclusions:
In short, the intelligent charging system of the automobile group belongs to the important project of implementing sustainable development, optimizing energy structure and promoting new energy development. The construction of intelligent charging system can meet the goal of energy structure upgrading, environmental protection, greenhouse gas control and energy conservation in China.